\documentclass{sf2a-conf2014}
\usepackage{graphicx}
\usepackage{hyperref}
\usepackage[]{natbib}
\usepackage{epstopdf}
\usepackage{amsmath,amsfonts,amssymb}

\def\BibTeX{{\rm B\kern-.05em{\sc i\kern-.025em b}\kern-.08em
    T\kern-.1667em\lower.7ex\hbox{E}\kern-.125emX}}
\bibpunct{(}{)}{;}{a}{}{,}  %%%%%%%%%%%%%  A&A bibliography style
\begin{document}

\TitreGlobal{SF2A 2014}

%%-----------------------------------------------------------------
%%      the top matter
%%

\title{Photophoresis in protoplanetary disks: a numerical approach}

\runningtitle{Photophoresis in PPD: a numerical approach}

\author{N. Cuello}\email{nicolas.cuello@univ-lyon1.fr}\address{Universit\'e de Lyon, Lyon, F-69003, France; Universit\'e Lyon 1, Observatoire de Lyon, 9 avenue Charles Andr\'e, Saint-Genis Laval, F-69230, France; CNRS, UMR 5574, Centre de Recherche Astrophysique de Lyon; Ecole Normale Sup\'erieure de Lyon, F-69007, France}
\author{F.\,C. Pignatale$^1$}
\author{J.-F. Gonzalez$^{1}$}

%% Keep this line, even if the page will be settled afterwards.
\setcounter{page}{237}

%%-----------------------------------------------------------------

\maketitle

%%-----------------------------------------------------------------
%%        The abstract
%% 
%%  Warning!  within the abstract:
%%  - do not use macros. 
%%  - do not use commands like: \cite, \citet, \citep ... etc.

\begin{abstract}
It is widely accepted that rocky planets form in the inner regions of protoplanetary disks (PPD) about 1 - 10 AU from the star. However, theoretical calculations show that when particles reach the size for which the radial migration is the fastest they tend to be accreted very efficiently by the star. This is known as the radial-drift barrier. We explore the photophoresis in the inner regions of PPD as a possible mechanism for preventing the accretion of solid bodies onto the star. Photophoresis is the thermal creep induced by the momentum exchange of an illuminated solid particle with the surrounding gas. Recent laboratory experiments predict that photophoresis would be able to stop the inward drift of macroscopic bodies (from 1 mm to 1 m in size). This extra force has been included in our two-fluid (gas+dust) SPH code in order to study its efficiency. We show that the conditions of pressure and temperature encountered in the inner regions of PPD result in strong dynamical effects on the dust particles due to photophoresis. Our simulations show that there is a radial and a vertical sorting of the dust grains according to their sizes and their intrinsic densities. Thus, our calculations support the fact that photophoresis is a mechanism which can have a strong effect on the morphology of the inner regions of PPD, ultimately affecting the fate of planetesimals.
\end{abstract}

%% Insert the keywords (to appear in the ADS indexing)
%% Keywords must be separated by a comma
\begin{keywords}
 protoplanetary disk, photophoresis, hydrodynamics, planet formation
 \end{keywords}

%%-----------------------------------------------------------------

\section{Introduction}
%%---------------------

Thousands of extra-solar planets have been detected in the past 20 years \citep{Batalha2013}, which undoubtedly shows that planets are common byproducts of stars. It is currently accepted that planets form in protoplanetary dusty disks around young stellar objects. This means that, throughout the evolution of the protoplanetary disk, dust grains have to grow from $\mu$m sizes to kilometric sizes to form planetesimals. However, the study of the motion of solids inside the disk leads to the so-called radial-drift barrier. This issue was first pointed out by \cite{W77} by solving the equation of motion for dust particles embedded in the gas disk of a PPD. A solid body orbiting a star in a keplerian fashion loses angular momentum through the interaction with the gas which orbits at a subkeplerian velocity. This is because the gas phase, which is assumed to be in hydrostatic equilibrium, is pressure supported while the dust phase is not. The difference between the orbital velocities of each phase can be interpreted as a headwind in the rest frame of the particle, which causes the particle to spiral down into the star.

Several mechanisms have been proposed to break the radial-drift barrier such as radial mixing \citep[and references therein]{KellerGail2004}, magnetic braking and dead zones \citep{Armitage2011}, particle traps \citep{Fouchet2010,Pinilla2012}, meridional circulation \citep{Fromang2011} and radiation pressure force \citep{Vinkovic2014}. However, it remains unclear if a single mechanism or a combination of them could actually prevent accretion. The aim of this work is to consider an extra mechanism called photophoresis, which has been first introduced by \cite{Rohatschek1995} through the study of illuminated particles in aerosols. \cite{Duermann2013} measured the strength of the photophoresis force on illuminated plates under similar temperature and pressure conditions that those found in PPD. Then, via an interpolation of their results, they computed the acceleration felt by solids of different sizes and porosities orbiting a 1 solar mass star. Plugging this extra force into the equations of motion for dust particles in a 1D model, they predicted that photophoresis might be able to stop the inward drift for large bodies ranging from a millimeter to a meter. Since the force is size dependent \citep{Duermann2013}, the photophoresis effects would ultimately lead to a radial sorting among different grain populations. However, to date, no studies have explored the complex dynamics obtained when considering this extra mechanism. This work is the first attempt to understand how photophoresis shapes the inner regions of PPD between 0.1 and 5 AU  through numerical simulations.

In section 2, we give a brief overview of the photophoresis force; in section 3, we explain how we model the PPD by means of  SPH simulations; in section 4, we present our results, and our conclusions are reported in section 5.

\section{Photophoretic force}
%%-------------------------
\cite{Rohatschek1995} derived the first semi-empirical model for the photophoretic force which connects low and high pressure regimes based on experiments and theoretical calculations. Indeed, there are three regimes for the photophoretic force, $F_{\rm ph}$, according to the pressure, $p$, of the gaseous environment around an illuminated particle: the low pressure regime for which $F_{\rm ph} \propto p$, the high pressure regime for which $F_{\rm ph} \propto 1/p$, and, in between, the transition regime when $F_{\rm ph}$ reaches its maximum. These regimes are determined by the value of a quantity called the Knudsen number defined as $\rm{Kn} = \lambda/a$, where $\lambda$ is the mean free path of the gas molecules and $a$ is the radius of the solid particle. Thus, $\rm{Kn} \gg 1$ for the low-pressure regime, whereas $\rm{Kn} \ll 1$ for the high-pressure regime. The semi-empirical expression for the photophoretic force reads as follows:
\begin{equation} \label{Fph}
 F_{\rm ph} = \frac{ 2 F_{\rm max}}{\frac{p}{p_{\rm max}}+\frac{p_{\rm max}}{p}} \, ,
\end{equation}
with
\begin{eqnarray}
 F_{\rm max} = \frac{a^2}{2} D \sqrt{\frac{\alpha}{2}} \frac{I}{k} \label{eq1} \, ,\\
 p_{\rm max} = \frac{3T}{\pi a} D \sqrt{\frac{2}{\alpha}} \label{eq2} \, , \\
 D = \frac{\pi \bar c \eta}{2T} \sqrt{\frac{\pi \kappa}{3}} \label{eq3} \, , \\
 \bar c = \sqrt{\frac{8RT}{\pi \mu}} \label{eq4} \, ,
\end{eqnarray}
where $I$ is the irradiance of the incident beam of light, $\alpha$ is the thermal accommodation coefficient (dimensionless and often taken equal to 1), $k$ is the thermal conductivity of the solid particle, $T$ is the gas temperature, $\eta$ is the viscosity of the gas, $\kappa$ is the thermal creep coefficient, which is equal to 1.14 \citep{Rohatschek1995}, $R$ is the universal gas constant and $\mu$ is the molar mass of the gas particle.

The photophoretic force can also be split into two components: the $\Delta T$-force mainly driven by the gradient of temperature between the illuminated and shadowed sides of the solid particle, and the $\Delta \alpha$-force which depends on the differences in composition of the particle. In this work, we consider homogeneous particles for which we vary the chemical composition and the size. Thus, we only need to compute the $\Delta \alpha$-force. Due to the lack of thermal conductivity data for large bodies, it is difficult to fix a value for $k$ which would depend on the temperature and the size. However, \cite{Opeil2012} and \cite{Loesche2012} showed, through measurements in chondrules and heat transfer calculations respectively, that a low porosity in dust aggregates suffices to lower thermal conductivities. For instance, the thermal conductivity for bulk silicates is of the order of magnitude of 1 $\rm{W\,m^{-1}\,K^{-1}}$ \citep{Opeil2012}, while if we consider the same silicates with a few percent of void, $k$ drops by at least one order of magnitude. This motivates our choice of a constant thermal conductivity $k=0.1\,\, \rm{W\,m^{-1}\,K^{-1}}$ as in \cite {Duermann2013}. All the other quantities on which equations \eqref{eq1} to \eqref{eq4} depend are functions of the disk local conditions and can be easily computed in our code as showed in the next section.

\section{Simulations}
%%-------------------------

In order to be able to compare our results to \cite {Duermann2013} we chose to study the same disk model, namely  the Minimum Mass Solar Nebula (MMSN) of \cite{H81}. We consider a protoplanetary disk of 0.013 $M_\odot$ mass with 1\% of dust by mass around a 1 $M_\odot$ star and a radial extension from 0.1 to 36 AU. The disk is vertically isothermal and $T(r) \propto r^{-q}$ with $q=1/2$. We focus on the inner regions from 0.1 to 5 AU since it is the range for which the photophoretic force is the more efficient \citep{Duermann2013}.

We adapt the code of \cite{Fouchet2005} to our study by including an extra term in the equation of motion for the dust particles. Gas and dust are considered as two separated fluids, which interact through aerodynamical drag. The code solves the equations of motion for each phase through the SPH formalism. \cite{Price2012} reviews the method and presents the recent developments. The equation of motion for the gas SPH particles reads:
\begin{equation}
 \frac{\rm d \boldsymbol{v_{a}}}{\rm dt} = \boldsymbol{P_{ab}} + \boldsymbol{M_{aj}} + \boldsymbol{D_{aj}} + \boldsymbol{G_{a}},
\end{equation}
which is the sum of the pressure term $P_{ab}$, the mixed pressure term $M_{aj}$, the drag term $D_{aj}$ and the gravity of the central star $G_{a}$. The subscripts $a$ and $b$ refer to gas particles and $i$ and $j$ for dust particles. For the dust particles the equation of motion is the following:
\begin{equation}
 \frac{\rm d \boldsymbol{v_{i}}}{\rm dt} = \boldsymbol{M_{ib}} + \boldsymbol{D_{ib}} + \boldsymbol{G_i}+\boldsymbol{F^{\rm photo}_{i}},
\end{equation}
where, in addition to the mixed pressure $M_{ib}$, the drag $D_{ib}$ and the gravity $G_{i}$ terms, we add the photophoresis force given by Equation \eqref{Fph}. It is important to note that the gas feels the pressure force whereas the dust does not, and that only the dust phase feels the photophoresis force. The photophoretic force depends upon the local properties of the gas at the position $\boldsymbol r$ of the dust particle: the temperature, the pressure, the viscosity and the amount of energy received from the star. In our simulations we consider that the medium is optically thin so that the radiant flux density at a given position $\boldsymbol r$ is simply computed as the luminosity over the surface of the sphere of radius $r$. This constitutes the main limitation of our calculations since photophoresis has no effect in optically thick regions. We are currently developing a more detailed model to include this effect.

In the simulations, we let a gaseous disk evolve from an initial surface density distribution given by $\Sigma \propto r^{-3/2} $. The initial velocity is keplerian, given by $v_k = \sqrt{G M_{\odot} /r}$. The reference values at 1 AU for the pressure and the temperature power laws are given by the MMSN model of \cite{H81}. Once the gas disk reaches equilibrium, we inject the dust particles on top of the gas particles with the same velocity. This state constitutes our initial state (Fig.~\ref{author1:fig1}-a). Then, we let the system evolve for an evolutionary time of 30 years approximately, \textit{i.e.} 30 orbits at 1 AU. We run a series of simulations to study the dependence of the photophoresis on the chemical composition (iron and silicates) and on the size of the solid bodies (1 cm, 10 cm and 1 m).

\section{Results}
\begin{figure}[ht!]
 \centering
 \includegraphics[width=0.8\textwidth,clip]{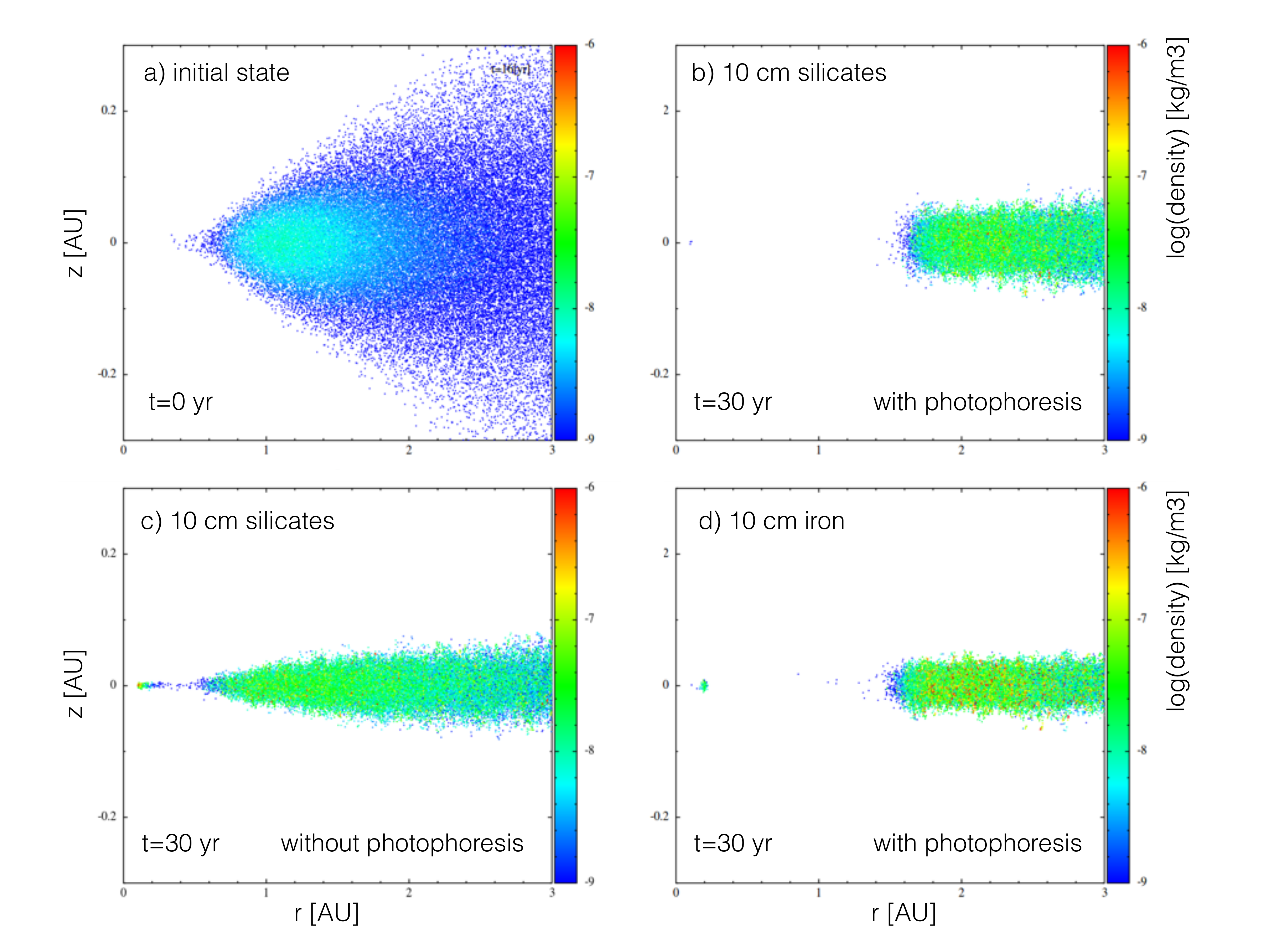}      
%% Note the ABSENCE of the extension .pdf , .eps or .ps  !
  \caption{Meridian plane cut of the dust distribution for the initial state of the dust phase (a) and the final state after 30 years of evolution for 10 cm silicate grains with photophoresis (b), without photophoresis (c) and with photophoresis for iron grains (d).}
  \label{author1:fig1}
\end{figure}

%%-------------------------
We show the initial state of the dust phase at $t=0$ (Fig.~\ref{author1:fig1}-a), the final state of the dust phase for 10 cm silicate grains with photophoresis (Fig.~\ref{author1:fig1}-b), for 10 cm silicate grains without photophoresis (Fig.~\ref{author1:fig1}-c) and for 10 cm iron grains with photophoresis (Fig.~\ref{author1:fig1}-d). We observe that there is a strong dust sedimentation as already observed in the simulations by \cite{Fouchet2005}. Radial migration is dramatically affected by the inclusion of photophoresis in the equations of motion. In fact, whithout photophoresis, the dust particles start to spiral down to the star and there is no outward motion. On the contrary, when we take photophoresis into account, the particles which are very close to the star between 0.1 and 1.8 AU tend to move outwards until they reach a stable orbit at around 1.8 AU for 10 cm grains. The location of the inner rim, \textit{i.e.} the radial migration, depends on the grain size: its position is at 0.5, 1.9, 1.5 AU for 1 cm, 10 cm and 1 m particles respectively. We see the strongest effect for 10 cm while the grains of 1 cm and 1 m are less affected by photophoresis. This result matches with the analytical calculation by \cite{Duermann2013}.

 Fig.~\ref{author1:fig1}-b and  Fig.~\ref{author1:fig1}-d show the different behavior for two different chemical compositions for a grain size of 10 cm: we notice that the vertical sedimentation is more efficient for iron grains than for silicates. This phenomenon is also observed in simulations without photophoresis since it solely depends on the intrinsic density ($\rho^{\rm sil}=3.2$ ${\rm g \, cm^{-3}}$ and $\rho^{\rm iron}=7.8$ ${\rm g \, cm^{-3}}$). Nevertheless, if we consider a more realistic disk made of a mixture of different species with a given distribution of sizes, then the sedimentation coupled with the different inner rim location patterns will have an effect on the mixing of solids in the inner regions of PPD. This will be the subject of a future work.

\section{Conclusions and future work}
%%--------------------
We have included photophoresis in our simulations in order to understand its effects on the inner regions of PPD. Even though our calculations present some limitations, our results show that photophoresis affects the structure of the inner rim of the dusty disk whereas the gaseous disk remains unchanged. Moreover, these preliminary results are in accordance with the predictions made by \cite{Duermann2013} concerning the different accumulation zones for different grain populations. Future work will explore the effect of this accumulation on the growth of large solids in the inner regions of PPD where we expect terrestrial planets to form. In fact, even if photophoresis is mainly effective for centimeter and meter sized bodies, it might lead to an efficient pile-up of particles close to the star. \cite{Chatterjee2014} recently proposed an inside-out planet formation scenario based on magneto-rotational instabilities (MRI). In this case, pebbles collect at the pressure maximum associated with the transition from a dead zone to an inner MRI-active zone. Alternatively, taking into account photophoresis effects,  solid bodies could accumulate and grow at the stable point defined by the transition between the optically thin and optically thick regions. In the optically thin part, particles move outwards since they mainly feel photophoresis, whereas in the optically thick one they move inwards due to the radial drift. If the dust-to-gas ratio is high enough in the accumulation zone, a planet could form at this location. Both mechanisms lead to systems with tightly-packed inner planets. It worths noticing that this planetary architecture seems to be one of the principal outcomes of planet formation since it has been detected in a large fraction of targets (more than 10\% of the stars) by the \textit{Kepler} mission \citep{Batalha2013}. 
 
% Optional acknowledgements
% -------------------------
\begin{acknowledgements}
This research was supported by the Programme National de Physique Stellaire, the Programme National de Plan\'etologie of CNRS/INSU. The authors are grateful to the LABEX Lyon Institute of Origins (ANR-10-LABX-0066) of the Universit\'e de Lyon for its financial support within the program "Investissements d'Avenir" (ANR-11-IDEX-0007) of the French government operated by the National Research Agency (ANR). All the computations were performed at the Service Commun de Calcul Intensif de l'Observatoire de Grenoble (SCCI). Figure~\ref{author1:fig1} was made with SPLASH \citep{Price2007}.
\end{acknowledgements}

%%-----------------------------
%%   Bibliography
%%-----------------------------
%%

%% The following lines are required when using BibTEX (strongly encouraged!):
\bibliographystyle{aa}  % A&A bibliography style file (aa.bst)
\bibliography{cuello} % your references in file: Yourfile.bib

\end{document}